\documentclass[aip,jap,amsmath,amssymb,reprint]{revtex4-1}
\usepackage{graphicx}

\begin{document}
\title{Normal metal - superconductor decoupling as a source of thermal fluctuation noise in transition-edge sensors } 

\author{K. M.  Kinnunen}
\author{M. R. J. Palosaari}
\author{I. J. Maasilta}


\affiliation{Nanoscience Center, Department of Physics, P. O. Box 35, FI-40014 University of Jyv\"askyl\"a, Finland}

\date{\today}
\begin{abstract}

We have studied the origin of excess noise in superconducting transition-edge sensors (TES) with several different detector designs. We show that most of the observed noise and complex impedance features can be explained by a thermal model consisting of three bodies. We suggest that one of the thermal blocks and the corresponding thermal fluctuation noise arises due to the high-frequency thermal decoupling of the normal and superconducting phase regions inside the TES film. Our results are also consistent with the prediction that in thin bilayer proximitized superconductors, the jump in heat capacity at the critical temperature is smaller than the universal BCS theory result.

\end{abstract}


\maketitle 

\section{Introduction}

A transition-edge sensor (TES) is a thin superconducting film that can be used as a sensitive thermometer when voltage biased within the normal metal - superconducting transition region and read out with superconducting SQUID sensors \cite{irwin}. TES based devices are used as extremely sensitive bolometers and calorimeters to detect radiation in a wide energy range from gamma-rays to sub-millimeter radiation \cite{enss}, and typically the thermal conductance to the bath is controlled by mounting the TES on a thin insulating SiN membrane. Although the performance of these detectors is already excellent,
the most sensitive TES devices have not yet reached the theoretical limits in energy resolution.
This is mostly due to excess noise that has been shown to be present in 
many devices \cite{ullom,NASA,hoevers,IM_LTD12,KK_LTD12}. Several candidates for the noise sources have been proposed, such as thermal fluctuations within the TES \cite{hoevers}, fluctuations in the Cooper-pair density \cite{FSN,secondFSN} or phase-slips \cite{slipsold,slipsnew}, but a definitive answer is still missing.

However, before resorting to more exotic noise sources to explain the data, one should be sure the known noise mechanisms are fully understood. These are the thermal fluctuations of the electrical degrees of freedom, or Johnson noise, and the thermal fluctuations of the energy degrees of freedom, usually called thermal fluctuation noise or phonon noise \cite{enss}. Both of these noise mechanisms are unavoidable in bolometric TES detectors; moreover, their magnitude depends on the details of the device in question. It is therefore of utmost importance to characterize the detector accurately, both electrically and thermally, before one can understand the origin of noise. 

A useful tool in characterizing TES detectors is the measurement of their frequency-dependent complex electrical impedance \cite{linde1}. Importantly, within the transition this impedance depends not only on the electrical, but also on the thermal circuit of the device, through the electrothermal feedback effect \cite{irwin,enss}. With the help of this technique, a more accurate picture of the electrical and thermal properties of TES sensors has emerged: First of all, it was realized that the dependence of the detector resistance on current, and not only on temperature, is critical for the detector response, as well \cite{linde1,enss}. This was later shown to influence the Johnson noise directly \cite{irwinnoise}, and some of the excess noise could then be explained as non-equilibrium Johnson noise. In addition, it has become clear that for many detectors, the simplest thermal circuit of one heat capacity connected to heat bath through one thermal conductance is not adequate \cite{NIST,NASASaab,nasaIEEE,cambridge,SRON,KK_LTD12,mikko}. A more complex thermal circuit then adds new components  to the thermal fluctuation part of the noise spectrum \cite{hoevers,galeazzi,enectali}.

Here, we present a study of the noise and complex impedance of several different designs of TES devices. Many of the devices are based on the so-called Corbino-geometry TES (CorTES)\cite{FSN, IM_LTD12}, where current spreads out radially from a central contact with radius $r_i$ into the outer contact at $r_o$, instead of flowing linearly [Fig. 1 (a)]. Although it is a bit more complicated to fabricate, it offers advantages in modelling, as the superconducting and normal regions separate due to the non-uniform current density profile. This means that we can determine the phase boundary radius $ r_b $ from the measured resistance $R$, which depends logarithmically on $r_b$ in Corbino geometry:

\begin{equation}
\label{eq:rb}
R \ln\left ( \frac{r_o}{r_i} \right )= R_{N} \ln\left ( \frac{r_b}{r_i} \right ) \Leftrightarrow \frac{r_b}{r_i} = \left ( \frac{r_o}{r_i} \right )^{R/R_{N}},
\end{equation}
where $R_{N}$ is the normal state resistance of the device. Once we know the size of the N and S phases, their expected theoretical heat capacities can be calculated as well. This level of theoretical description is not possible in more common square shaped TES devices.

CorTES has also shown an excess noise component, if compared to the simplest thermal model. This was originally explained by the fluctuations of the N-S boundary, or fluctuation superconductivity noise (FSN) \cite{FSN}. However, here we show that most of the "excess" noise is simply internal thermal fluctuation noise originating from a more complex thermal circuit within the device. By analyzing the data thoroughly, we suggest that most of the "excess" noise is generated by the thermal decoupling of the spatially separated superconducting and normal regions of the devices. Data from typical square shaped devices also supports this picture: Due to non-uniformities of real devices and/or the lateral proximity effect \cite{latp}, phase separation can still take place, although not in such a controllable manner as in CorTES devices.

\begin{figure}
\includegraphics{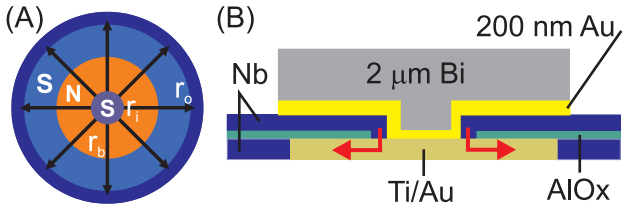}
\caption{(Color online) (a) Diagram illustrating the radial current distribution and separation into normal (N) and superconducting (S) phases in a CorTES. $ r_i $ and $ r_o $ are the radii of the inner and outer superconducting contacts and $ r_b $ is the phase boundary radius. (b) A schematic side view of a CorTES with an absorber. The arrows indicate the path of the bias current. \label{CorTES}}
\end{figure}

\section{Thermal modelling}

\begin{figure}
\includegraphics{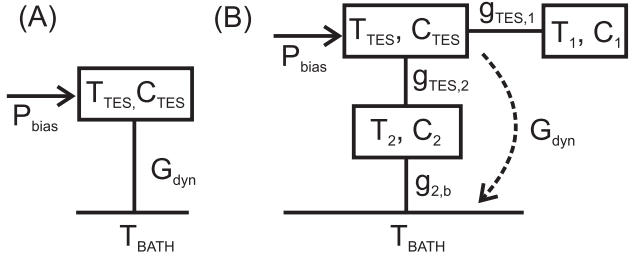}
\caption{(a) Basic calorimeter thermal model. (b) Three-body model used in this work. \label{block}}
\end{figure}

\begin{figure}
\includegraphics{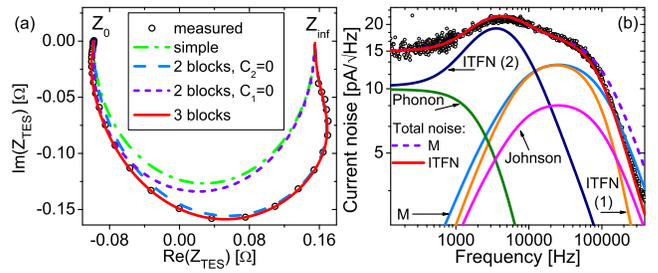}
\caption{(Color online) (a) A comparison of one- two- and three-body fits to a typical impedance data. (b) An example of a current noise spectrum from a CorTES device, showing the minor difference between fitting the high frequency noise with ITFN or with excess Johnson (M) noise.  \label{MvsITFN}}
\end{figure}

The simplest thermal model of a TES device is shown in Fig. \ref{block}(a), consisting of a single heat capacity connected to the heat bath. The complex impedance $Z$ of a device with this one-body thermal circuit always traces a semicircle in the complex plane as a function of frequency \cite{enss}. Thus, it is experimentally straightforward to determine whether a more complex thermal circuit is required by studying the shape of the $Z$ curve. Whenever additional thermal blocks are added to the system, $Z$ develops bulges outwards from the ideal case ( see Fig. \ref{MvsITFN} (a)). The shape and size of the new features in $Z$ depend on the heat capacities $C_i$ of the added bodies and the thermal conductance links $g_i$ between them, in such a way that the features grow in size with both $C_i$ and $1/g_i$. 

Fig. \ref{MvsITFN} (a) shows a typical measured impedance curve of a CorTES device, together with a one-, two- and three-body theoretical model fits. Although simpler one and two-body models seem to fit some device designs by other groups \cite{iyomoto, NASASaab, SRON, Eckart}, for our devices a three-body model is the simplest model that fits the data well, as is clear from Fig. \ref{MvsITFN} (a). Recently, simple three-body models have also been used by other groups for more accurate modelling of their devices \cite{NIST,nasaIEEE}. 

In this work, we have therefore used the thermal model shown in Fig. \ref{block}(b), which is the simplest three-body model that is physically justifiable for our devices. In addition to the TES film with heat capacity $C_{TES}$ where the bias power $P_{bias}$ is dissipated, there is a hanging thermal body with heat capacity $C_1$, and an intermediate heat capacity $C_2$ between the TES and the heat bath.
$ C_2 $ could be associated with the supporting SiN membrane, and $C_1$ could represent the absorber on top of a TES film [Fig. 1 (b)], but in general it could be some other part of the device also. Most devices studied here do not actually have absorbers. We call this the IH (intermediate + hanging) model, and use analytical expressions for the impedance and noise of a TES with this thermal circuit presented in \cite{mikko}. Their derivation and full theoretical discussion will be published elsewhere \cite{TPB}. We simply note that with each new thermal block, a new thermal noise source is also introduced to the system, arising from fluctuations in energy between the bodies through the thermal link. We call the noise due to any additional block internal thermal fluctuation noise (ITFN) to distinguish it from the thermal noise between the system and the heat bath (phonon noise).

\subsection{Intermediate body}

The intermediate body is characterized by its temperature $T_2$ and heat capacity $C_2$. We emphasize that we do not make the simplifying assumptions that $ T_2 = T_{TES} $ or that $ T_2 = T_{bath}$, when the TES is biased. Associated with the intermediate block, we thus have four different values of dynamic thermal conductance, two for each physical link connected to $C_2$, evaluated at each temperature end of the link \cite{TPB}. 
The parameters $ G_{dyn}=dP/dT_{TES} $ and $ T_{TES} $ can be calculated from a careful analysis of the measured I-V characteristics, as described in Ref \cite{linde2}. 

In our fitting procedure we use $T_2$, $C_2$ and $g_{TES,2}(T_{TES})$ as the free parameters. The other unknown conductances $g_{TES,2}(T_2)$, $g_{2,b}(T_{bath})$ and $g_{2,b}(T_2)$ are then calculated using the free parameters and the known values of $T_{TES}$ and $G_{dyn}$,
as \cite{TPB}
\begin{equation}
 G_{dyn}=\frac{g_{TES,2}(T_{TES})g_{2,b}(T_2)}{g_{TES,2}(T_{2})+g_{2,b}(T_2)}.
\end{equation}  
To simplify the model and to minimize fitting parameters, it is also assumed that the links on both sides of the intermediate block have the same thermal exponent $n$, i.e. 
$g_{TES,2}(T_j) = AT_{j}^{n-1}$ and $g_{2,b}(T_j) = BT_{j}^{n-1}$. This is physically reasonable if all the conductances are dominated by the phononic transport properties of the SiN membrane.  

Later, we will show that the intermediate body fitting parameters are nearly constant throughout the transition, and only change when $ T_{bath} $ is changed.  Also, $C_2$ appears to depend on both the area and thickness of the membrane. This behavior is thus consistent with the assumption that the intermediate body represents the SiN membrane phonons. Furthermore, $ T_2 $ is usually always slightly below $ T_{TES} $, and not close to the bath temperature.

\subsection{Hanging body}

In contrast to the intermediate body, the second added heat capacity $C_1$ in our model is hanging [Fig. \ref{block} (b)], which means that no steady state power flows through it, and the average values of $T_{TES}$ and $T_1$ are equal. This simplifies the description in comparison to the intermediate body, as only one thermal conductance $g_{TES,1}$ is required. The Joule power of the TES bias current is all assumed to be dissipated inside $C_{TES}$. 
In real X-ray and $\gamma$-ray devices with thick absorbers on top of the TES film \cite{iyomoto,NIST,mikko}, $C_1$ could well be the absorber, in which case $g_{TES,1}$ describes the thermal conduction within the TES and the absorber. In devices without an absorber (most samples in this work), the idea is to model electronic degrees of freedom, thus $g_{TES,1}$ models the thermal conduction within the TES film. 

The hanging block is naturally still a simplification, in reality there can be also a thermal coupling from $C_{1}$ directly to the bath or to $C_{2}$. However, the effect of the missing link, which would have a magnitude of the order of $g_{TES,2}$, is believed to be minor for the devices studied here, where $g_{TES,1}$ is much larger than $g_{TES,2}$, as will be seen later. 
Any effects of temperature gradients within the TES film are also not captured in this simple model. Numerical estimates of the gradients within a CorTES device \cite{AL} have shown that the isothermal simplification is a reasonable assumption.    

For the CorTES devices with phase separated N and S regions, one may at this point wonder, if the blocks $C_{TES}$ and $C_{1}$ could actually directly represent the normal and superconducting regions of the TES film, respectively. The great benefit of the CorTES geometry  is that we can a priori calculate the values of $C_{TES}$ and $C_{1}$ as a function of bias, which is not possible for the more standard square shaped devices. As we later show, the fitted values of the heat capacities and their behavior as a function of bias point do indeed follow a trend predicted by this interpretation. $g_{TES,1}$ is thus still the electronic thermal conductance, but now it can be affected by the location of the NS boundary, for example. Understanding of how $g_{TES,1}$ should behave as a function of the bias (phase boundary location) is quite sketchy at the moment, and it needs to be studied further theoretically.   

The hanging body $C_{1}$ produces internal thermal fluctuation noise (ITFN) that appears in current noise spectra at the high-frequency side of the effective thermal time constant, just like Johnson noise \cite{mikko,TPB}, see Fig. \ref{MvsITFN} (b). In addition, when $ g_{TES,1} $ is large enough, the roll-off for the ITFN noise occurs at very high frequencies, even above the electrical cut-off frequency of the read-out circuit. Thus, the ITFN noise produced by the hanging body can look very similar to the Johnson noise. Quite often in the past, excess noise in TES devices has been quantified by the parameter $M$ \cite{ullom,IM_LTD12,SRON}, which is a measure of excess Johnson noise. Thus, one has to be very careful about drawing conclusions based on just the noise data: If the thermal circuit is not adequately characterized, one can misinterpret ITFN noise as excess Johnson noise. In our devices the time constant for ITFN roll-off is indeed near the electric cut-off of our read-out circuit, so that fitting the spectra either with the three-body ITFN model, or two-body + excess Johnson noise ($M$ parameter) give almost identical results, as illustrated in Fig. \ref{MvsITFN} (b). The only difference seen is that with the ITFN model, the roll-off has a steeper slope, as it is a higher order roll-off produced by the combination of the thermal and electrical circuits. However, the point is that in many cases the impedance can only be fitted with the three block model [see Fig. \ref{MvsITFN} (a)], which supports the choice of ITFN noise over other noise sources.


\section{ Experimental setup and methods}
\label{setup}

The experiments were performed in a compact homemade dilution refrigerator with a base temperature $T= 40$ mK. The TES devices were voltage biased by a shunt resistor (8.9 m$\Omega$) at the sample stage, and a NIST two-stage SQUID \cite{cherv} mounted on the 1 K stage (equivalent input current noise density $\sim 4$ pA/$\sqrt{\mathrm{Hz}}$ and geometric input inductance of 300 nH) served as the current amplifier for the readout. The SQUID was operated in the flux locked loop, with dedicated room temperature electronics designed by SRON. More details on the setup can be found in \cite{Kimmothesis}. 

All studied TES devices were first characterized by measuring their resistance vs. temperature transitions with a four-probe lock-in measurement, to find the normal state resistance $ R_N $ and critical temperature $ T_c$. Next, each detector was connected to the SQUID readout, and a series of current-voltage (I-V) curves was measured at several different bath temperatures.  The thermal exponent $n$ was extracted from the $T_{bath}$ dependence of the I-V data, as explained in Ref. \cite{linde2}. Then, using the known $n$, the TES temperature $T_{TES}$  and dynamic thermal conductance $G_{dyn}$ at any point in the transition were determined from the I-V curves \cite{Kimmothesis}. We also calculated the transition steepness parameter $ \alpha_{TOT,IV} = (T/R)dR/dT $, which is a dimensionless measure of the sensitivity of the TES \cite{irwin}.

Finally, a set of noise and complex impedance measurements were performed at a desired $ T_{bath} $ and at various bias points within the transition. The TES impedance was extracted from the measured circuit impedance by dividing out the effect of the read-out circuit transfer function, as described in ref. \cite{transfer}. From the TES impedance curves, the low and high frequency limits $ Z_0 $ and $ Z_{\infty} $ can be estimated, which can then be used to calculate the important transition parameters $ \alpha = (T/R)\partial R/\partial T|_{I_0} $ and $ \beta = (I/R)\partial R/\partial I|_{T_0}$ as  

\begin{equation}
\label{eq:alpha}
\alpha = \frac{n}{\phi}\frac{Z_0 - Z_{\infty}}{Z_0 + R_0}, \; \beta = \frac{Z_{\infty}}{R_0} - 1
\end{equation}
where $ \phi = 1 - (T_{bath}/T_{TES})^n $ and $ R_0 $ is the TES resistance. Additionally, $ \alpha _{TOT,Z} = (n/\phi)(Z_0 - R_0)/(Z_0 + R_0)=[2\alpha+(n/\phi)\beta]/(2+\beta)$ from the impedance data should give \cite{AIP} the same value as $ \alpha _{TOT,IV} $, so that we can check for consistency between the impedance and I-V measurements. Notice how $\alpha _{TOT}$ depends both on $\alpha$ and $\beta$.   

We have performed impedance measurements between 4 Hz and 100 kHz both by the white noise method \cite{linde1}, where a white noise excitation is used and the broadband response is measured, and by the lock-in method where a sinusoidal excitation is used at certain frequency points \cite{transfer}. 
Both methods can in general be used, however, we have seen that the sine-wave method is more reliable, as the total heating power generated by the excitation is typically less in the sine-wave method.  Data with heating problems could be identified by the consistency check mentioned above: With excess heating $ \alpha _{TOT,Z} $ did not agree with $ \alpha _{TOT,IV} $, indicating a heating induced shift in the bias point. 

The measured impedance and noise data are finally fitted {\em simultaneously} to the IH model  equations. We emphasize that the fitting is done by eye, and free fit parameters are varied manually, as high-dimensional non-linear least-squares fitting would be demanding to implement. In some cases we found that the three-body equations still did not explain all the observed noise, even if the impedance fit was good. In those cases we have quantified the remaining truly excess noise as excess Johnson noise using the M-parameter. Note also that although here we only show the complex plane plots of $Z$, we also make sure that the real and imaginary parts fit separately as a function of frequency. Included in the fits, but not shown in the plots because of their small values, are the Johnson noise of the known shunt resistor, and the equivalent white input noise of the SQUID. We have also used the equation for the (lowest order) non-equilibrium TES Johnson noise \cite{irwinnoise} $V_n^2=4k_BTR_0(1+2\beta)$  in the analysis for all the devices. Even though $Z$ is measured only up to 100 kHz, the theory curves are always calculated up to 2 MHz, in order to see the high-frequency differences between the fits, as demonstrated in section \ref{sec:slice1}.

\section{Experimental results}

In this section we present the measured data and fit results from several different detectors. Table \ref{TEStable} lists some of the key parameters for each device studied. Most of them have radially spreading current distribution as in the fully circulary symmetric CorTES device shown in Fig. \ref{CorTES}. However, the device labelled STES is divided into four equal size parallel slices (see inset, Fig. \ref{STESparam}), and the detectors labelled Slice 1 and Slice 2 are individual slices (Fig. \ref{111fits}). Also, data from one traditional square-shaped device is also presented. All measurements were done at a regulated bath temperature of 60 mK unless stated otherwise. The detectors were not shielded against external magnetic fields, nor was earth's field compensated for. 
 \begin{table}[h]
 \caption{\label{TEStable} Device parameters for the TESs measured in this work.}
 \begin{tabular}{r c c c c}
TES & Ti/Au [nm] & $T_c$ [mK] & $R_N$ [m$\mathrm{\Omega}$] & SiN [$\mathrm{\mu m}$] \\ \hline
CorTES & 40/55 & 98 & 200 & 0.30$\times$800$\times$750 \\
STES & 40/55 & 99 & 220 & 0.30$\times$800$\times$750 \\
Slice 1 & 71/105 & 126 & 166 & 0.30$\times$830$\times$730 \\
Slice 2 & 58/83 & 162 & 220 & 0.75$\times$830$\times$730 \\
Square & 48/70 & 156 & 425 & 0.75$\times$460$\times$410 \\
 \end{tabular}
 \end{table}

\subsection{CorTES}
\label{sec:cortesdata}

\begin{figure}
\includegraphics{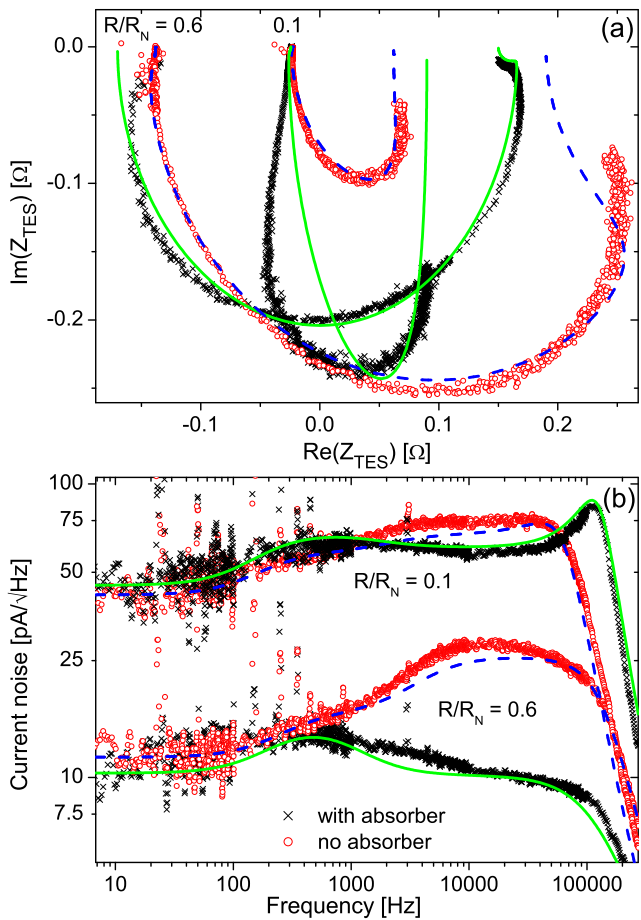}
\caption{(Color online) Comparison of (a) the impedance (measured with the noise method) and (b) the noise of a full CorTES with (black crosses) and without (red open circles) an absorber, at two different bias points. The lines are best fits to the data using the IH model. No $M$-parameter was used. \label{OTESdata}}
\end{figure}

The data we have measured on a full CorTES turned out to be quite difficult to fit accurately even with the three-body IH model, as seen from the examples shown in Fig. \ref{OTESdata}. The general trends are, nevertheless quite well reproduced, including the very strong deviations of $Z$ from the simplest circular shape and all the trends in the noise spectra. Typical values for the  most important fit parameters were $C_{1} \sim 0.2$ pJ/K, $C_{2} \sim 0.35$ pJ/K, $C_{TES} \sim 0.05$ pJ/K and $g_{TES,1} \sim 20-30$ nW/K throughout the transition for the bare devices without an absorber. Fig. \ref{OTESdata} also shows the data after the deposition of a 2 $\mu$m thick Bi absorber on top, as shown schematically in Fig. \ref{CorTES}(b). In that case, all other parameters stayed about the same, except for $C_{1}$, which, quite reasonably, doubled to $\sim 0.4$ pJ/K. We should note that the obtained $C$ values are reasonable if compared to estimates from the detector size, except for $C_{2}$ which is surprisingly large. This could be because of the 120 nm thick AlOx insulator layer separating the Nb bias lines [\ref{CorTES}(b)] in the CorTES device, creating unwanted thermal links and heat capacity inside, as shown in Ref. \cite{KK_LTD14} for a different device.   

Very accurate determination of how the addition of the absorber affected the detector characteristics is unfortunately complicated, because the absorber processing changed also the transition properties, by lowering the $ T_c $ from 98 mK to 85 mK, and by broadening the transition. For the bias values shown in Fig. \ref{OTESdata}, $\alpha _{TOT,IV}$ decreased from  $\sim 300$ to $\sim 50$ at $R/R_{N}=0.6$, whereas at $R/R_{N}=0.1$ it stayed constant at $\sim 150$. These changes influence the detector responsivity, which means that noise and $Z$ are affected not only through the increase of $C_{1}$ but also directly. The large difference of noise at $R/R_{N}=0.6$ can be attributed mostly to the change in $\alpha$, whereas at $R/R_{N} = 0.1$ the lowering of noise is consistent with the lower $T_c$ of the device after the addition of the absorber.

\subsection{STES}

\begin{figure}[h]
\includegraphics{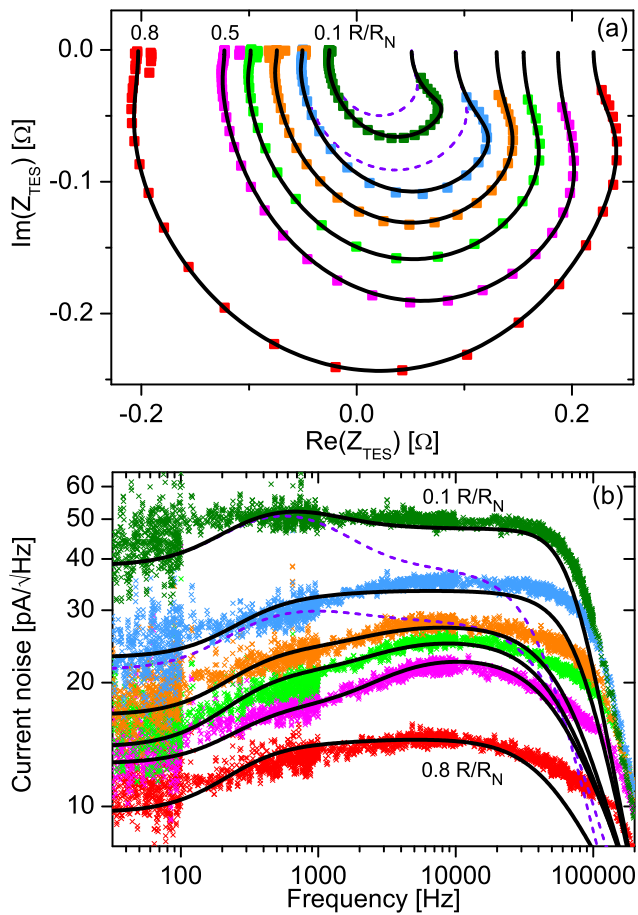}
\caption{(Color online) Measured (symbols) and fitted (lines, IH model) (a) impedance (measured using the lock-in method up to 100 kHz) and (b) noise in the STES device. Decreasing bias makes the impedance curve smaller and noise level higher. Bias values shown are $R/R_{N}=0.1$, 0.2, 0.3, 0.4, 0.5, 0.8, rest are omitted for clarity. Dashed lines show fits with alternative parameter values shown in Fig. \ref{STESparam} as crosses. \label{STESfits}}
\end{figure}

\begin{figure}[h]
\includegraphics{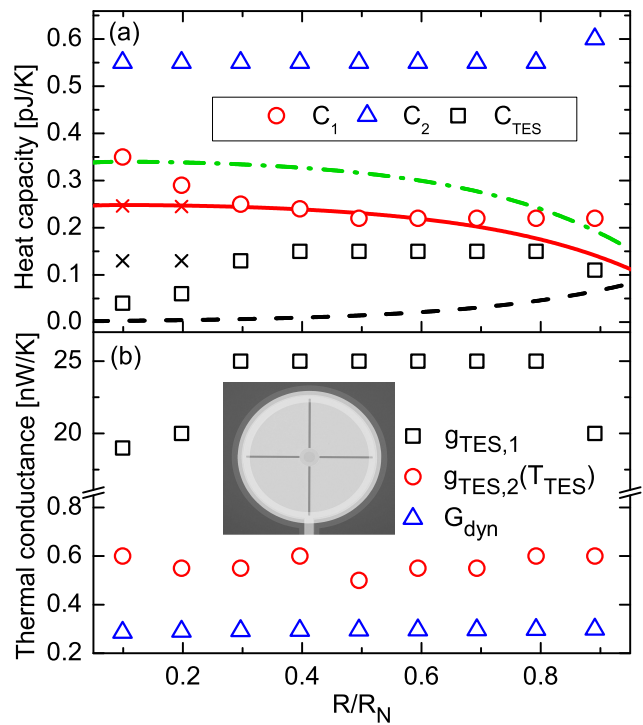}
\caption{(Color online) Parameters obtained from the fits to the data in Fig. \ref{STESfits} as a function of the bias point.(a) The heat capacities, and (b) the thermal conductances. $G_{dyn}$ is fixed by the I-V data and is not a fit parameter. Theory curves show the expected dependence of the normal region (dashed line) and the superconducting region with two different models for the jump at $T_c$: full jump (dash-dotted) and suppressed according to \cite{kozorezov} (solid line). Crosses show alternate parameter values used for "bad fits" shown in Fig. \ref{STESfits}. Inset in (b) is an SEM image of the STES. \label{STESparam}}
\end{figure}

We have also measured a set of data from another absorberless CorTES device, shown in Fig. \ref{STESfits}. It was otherwise nominally identical to the full device discussed above, but was cut into four parallel slices (dubbed STES), with an SEM image shown in the inset of Fig. \ref{STESparam}(b). In this case, the impedance fits to the IH model look nearly perfect, and the noise data agrees also quite well without the need of an empirical $M$ parameter, except at the very highest frequencies near the cut-off (reason unknown at the moment). Achieving better fits than in the full CorTES case was easier, most likely because the STES device has a lower responsivity (lower maximum $\alpha \sim 200$ ).  In Fig. \ref{STESparam} we plot the obtained fit parameters of interest as a function of the bias point $R/R_{N}$. The heat capacities, Fig. \ref{STESparam} (a), do not vary strongly, and their values are generally consistent with the CorTES results. $C_{2} \sim 0.55$ pJ/K is again high, even higher than in the CorTES device. To compare the data with the suggestion that $C_{TES}$ and $C_{1}$ are associated with the normal and superconducting regions, respectively, we also show theoretical curves corresponding to the calculated heat capacities of the normal and superconducting phase regions, using the location of the phase boundary calculated from Eq. \ref{eq:rb}, the known film thicknesses and Sommerfeld constants $\gamma_{Ti}= 330$ J/K$^2$m$^3$ and $\gamma_{Au}= 65$ J/K$^2$m$^3$ from literature \cite{constants}. In addition, the heat capacity of the S region was calculated using two different assumptions: (a) the jump at $T_c$ is given by simple BCS theory as $1.43 C_{N}$ (dash-dotted line), and (b) the jump is slightly suppressed due to proximity effect according to Ref. \cite{kozorezov}. We see that $C_{1}$ is clearly consistent with it being associated with the S region, however $C_{TES}$ is a bit elevated compared to the simplest theory. Moreover, lower in the transition at $R/R_{N} < 0.3$, there seems to be a trend that $ C_{1}$ increases and $ C_{TES}  $ decreases. This trend is very clear in the data, as forcing the values of $C_{1}$ and  $C_{TES}$ for  $R/R_{N} < 0.3$ to equal the values at $R/R_{N} = 0.3$ produces very poor fits, as shown in Figs. \ref{STESfits} and \ref{STESparam}.  

\subsection{Slice TES}

The usual CorTES devices have an AlOx insulator layer between the bias leads. Furthermore, from the high values of $C_{2}$ obtained above, and from previous measurements in Ref. \cite{KK_LTD14}, we suspect that the AlOx layer could be responsible for the high value of $C_{2}$. Therefore, we also studied devices where only a quarter "slice" of the full CorTES disk is retained: this way the insulator layer is not required, as the center contact lead can come from the opposite side (see inset Fig. \ref{111fits}). However,  we still retain a geometry that promotes a phase separation.   Here we report on measurements on two such samples, with different Au/Ti thicknesses, leading to different $T_c$ and $R_{N}$ values (Table \ref{TEStable}). The Au/Ti layer thicknesses were increased compared to the full CorTES devices, in order to keep $ R_N $ approximately the same.  Also, the thickness of the SiN membrane (300 nm) was chosen to be the same as in the usual CorTES devices for sample 1, whereas sample 2 had much thicker SiN (750 nm).  

\subsubsection{Sample 1}
\label{sec:slice1}
\begin{figure}[h]
\includegraphics{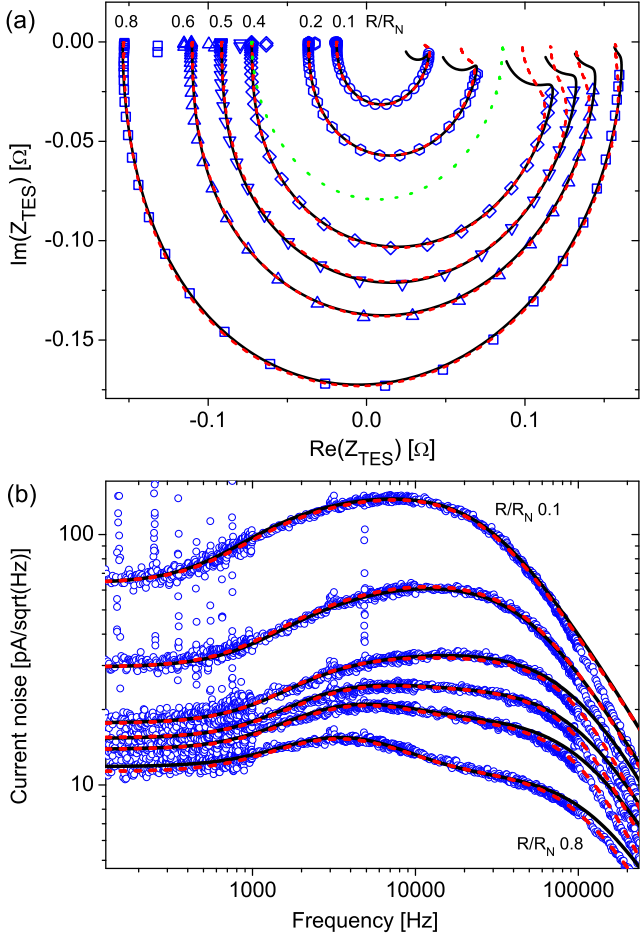}
\caption{(Color online) (a) Impedance and (b) noise data for the slice TES, Sample 1 (symbols), for $R/R_{N}$=0.1, 0.2, 0.4, 0.5, 0.6, 0.8, rest omitted for clarity. We also show the best fit for case 1 (solid black line) and case 2 (dashed red line).  The dotted green line in (a) is the impedance according to the thermal model of Fig. \ref{block}(a), illustrating how much our data deviates from the simple one-body case, and it was calculated using $Z_{0}$ and $Z_{\infty}$ obtained from case 1 at $R/R_N$ = 0.4.  \label{112fits}}
\end{figure}

The first slice TES has relatively low values of $ \alpha \sim 60 - 80 $, with the result that the measured impedance data, shown in Fig. \ref{112fits}, lacks any striking features. The data is therefore too easy to fit: That is, we can find several ways to fit the data (both impedance and noise), and choosing the best fit is not simple based on this data alone. To illustrate this, we have fitted the same data in three different ways, with the resulting fitted curves almost identical up to 100 kHz, but with significant differences in the obtained parameter values. The three cases we consider are: Case 1: we force $ C_{TES} $ to the theoretically calculated value and let the other parameters vary freely. In case 2,  $C_{TES} $ is not fixed but also free (as in all the fits for the CorTES and STES devices above). Finally, in case 3, we set $ C_{1} = 0 $, reducing the thermal model to two bodies: the TES and the intermediate. 

From the CorTES sample results above, we still expect that the three-block model is required to describe the sample physics, thus we first discuss the comparison between cases 1 and 2, and only later comment on the case 3 fits. Fig. \ref{112fits} shows  a comparison of cases 1 and 2 with the data.  It is clear that both cases fit the data well, and deviate from each other only at high frequencies close to 100 kHz, with case 1 making a more pronounced kink in the calculated impedance and slightly weaker roll-off for the noise.  

\begin{figure}[h]
\includegraphics{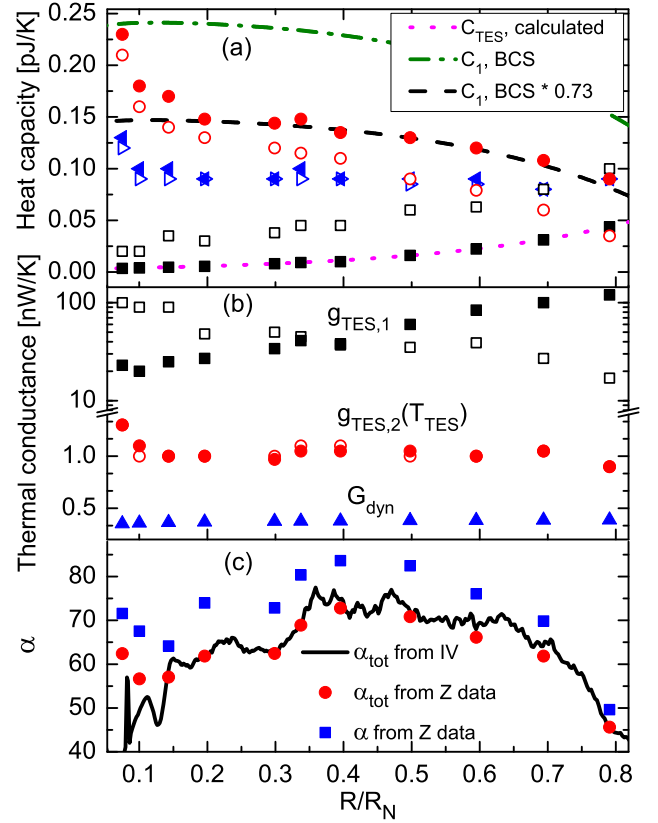}
\caption{(Color online) Parameters obtained from the fits in Fig. \ref{112fits}. Case 1 is shown by filled symbols, and case 2 with open symbols. (a) Squares: $ C_{TES}$, triangles: $ C_{2} $ and circles: $ C_{1}$. Lines are the theoretical calculations as discussed in the text. (b) Thermal conductances of interest. (c) $ \alpha $ and $\alpha_{TOT}$ parameters calculated from the I-V (line) and impedance data (points).  \label{112param}}
\end{figure}

Looking at the obtained parameters plotted in Fig. \ref{112param}, we find case 1 quite interesting.  With $ C_{TES} $ forced to the theoretical values, $ C_{1} $ follows the calculated theoretical value of the reduced BCS heat capacity \cite{kozorezov}, supporting again the proposed picture of thermal decoupling between the N and S phases. As expected, due to the lack of AlOx, $C_{2} \sim 0.1$ pJ/K is much reduced compared to the CorTES and STES samples, even though the SiN membrane dimensions are the same. Case 2 reproduces all the trends, but now $C_{TES}$ is higher than the theory (as in the STES device) and $ C_{1} $ lower, but with their sum approximately constant between the two cases. We also note, how for both cases $ C_{1} $ grows fast below $R/R_{N} < 0.2$, similar to the STES device.   

On the other hand, comparing the two cases in terms of $ g_{TES,1} $ fit values produces a striking difference: The trends are quite clear, but opposite for the two cases. Thus, it is hard to draw solid conclusions, except that one can see that the highest values obtained $\sim 100$ nW/K are consistent with the increased TES bilayer film thickness, if compared to the STES results. The trend in case 1 of decreasing $ g_{TES,1} $ with decreasing $R$ is of course consistent with the fact that thermal conductance starts to decrease in the superconducting phase, and that the size of the S phase increases when going down in bias.  

In cases 1 and 2 we did not need to use the $M$-parameter (excess noise) in the upper part of the transition, but had to include it in the lower part,  in case 1 below 30 \% bias and in case 2 already at 50 \%. This is in contrast to the CorTEs and STES fits, where $M$-parameter was not used at all (those fits were not as accurate, though).  The fitted values of M are shown in Fig. \ref{MbetaCtot}(a). For this sample, we also found that in a narrow bias range $ R/R_N$ = 0.3 to 0.35 there was an  enhancement of the mid-frequency (1 kHz - 30 kHz) noise that could be explained neither by the thermal circuit nor an $M$-noise component. Curiously, this bias range   
 corresponds to the sudden drop in the $\alpha$ parameters, as can be seen in Fig. \ref{112param}(c)).

\begin{figure}[h]
\includegraphics{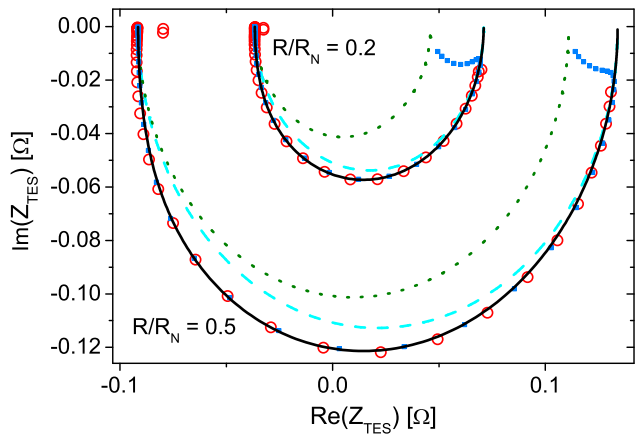}

\caption{(Color online) Impedance fits in cases 3 (solid black line) and 1 (blue squares) compared. Red circles are the measured data. For comparison with the simple calorimeter model of Fig. \ref{block}(a), the dashed lines show the calculated one-block impedances using $Z_{\infty}$ of case 1 (short dash) and case 3 (long dash).
\label{112Cabs0}}
\end{figure}

To study Case 3, we show in Fig. \ref{112Cabs0} representative impedance fits at two bias points (solid lines), with case 1 fits included for comparison. Again, if we only look at frequencies up to the measurement maximum, we do not see any difference in the fits. The difference in the noise fits (not shown) is also limited to a small variation in the roll-off. In case 3 we naturally do not have the high frequency ITFN component present because $ C_{1} $ is missing, and thus a large part of the noise at high freqencies has to be accounted for by excess noise with $M$ parameter, at all bias points. The effect of the analysis case on the obtained $M$-parameter values vs the bias point is shown in Fig. \ref{MbetaCtot} (a). In addition,  the chosen fitting case has a big effect also on $ \beta $, as it is proportional to $ Z_{\infty} $ [Fig. \ref{MbetaCtot} (b)].  

The plots remind us that many parameters strongly depend on the chosen thermal model and how it is interpreted. If we do not know with certainty which of the possible thermal models is correct, we should not jump to conclusions about the nature of the excess noise based on the dependence of $M$ parameter on bias, magnetic field etc. In other words, the $Z$ data may look like it fits a simpler one- or two-block model, but the underlying true model could still be different. For example, if one is developing a detector where $ \beta $ is an important parameter, mistakes can be made in TES design if the decisions are based on $ \beta $ from a wrong fit. For the case of this particular device, we believe that Cases 1 and 2 are more accurate because of the evidence from the other samples studied here, where two-block model fits are impossible to the impedance data.    

\begin{figure}
\includegraphics{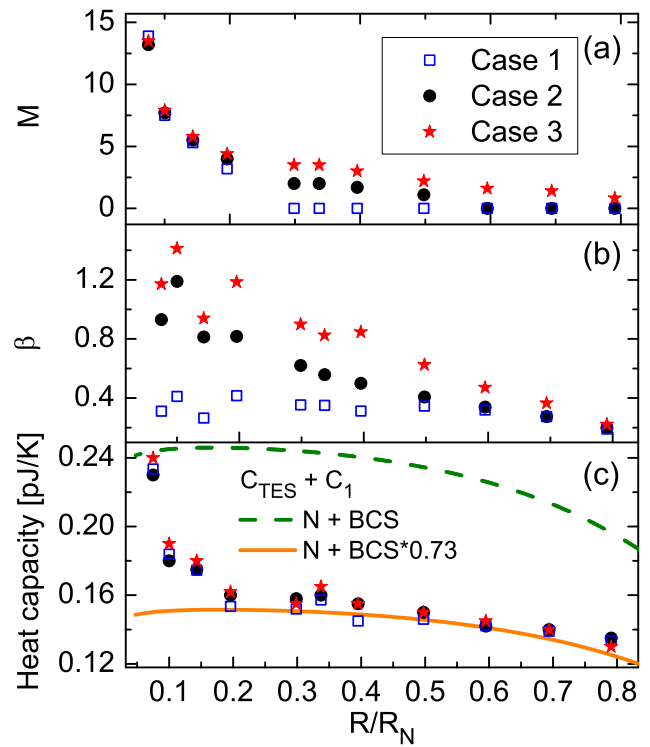}
\caption{(Color online) Comparison of (a) M, (b) $ \beta $ and (c) the sum of $ C_{TES} $ and $ C_{1} $ between different fits cases. Symbols: values from fits. Line: BCS heat capacity with the bilayer correction, dashed line, full BCS theory. \label{MbetaCtot}}
\end{figure}

Another interesting point to notice is that even though we have three different sets of values for $ C_{TES} $ and $ C_{1} $, the sums $ C_{TES}+C_{1} $ are essentially constant between the different fitting cases,  as shown in Fig. \ref{MbetaCtot}(c). The values are also consistent with the modified BCS theory \cite{kozorezov}, except at the lowest bias points, where the rapid increase mentioned before is clearly seen.  

At this point we wish to speculate about the observations at low bias points. A possible scenario for the effects seen (increasing $C$, increasing $M$-noise) could well be related to vortex physics. If vortices are not pinned, they could contribute to heat capacity and also generate excess noise by the so called phase-slip shot noise mechanism \cite{slipsold,slipsnew}. The sudden drop in $ \alpha_{TOT,IV} $ is also correlated with the onset of the need for $M$- parameter in Case 1. The lower $\alpha$ would then correspond to extra resistance in the transition, caused by the vortex motion.  In this light we can perhaps also understand why the full CorTES devices do not require $M$ to fit their noise, as the extra Nb top layer over the TES film may pin vortices more strongly, and prevent the generation of excess phase slip noise. Another option is that some of the excess $M$-noise could be generated by the FSN mechanism \cite{FSN}. However, FSN theory in its current state cannot explain the increased heat capacity or lowered $\alpha$. 

\subsubsection{Sample 2}

\begin{figure}
\includegraphics{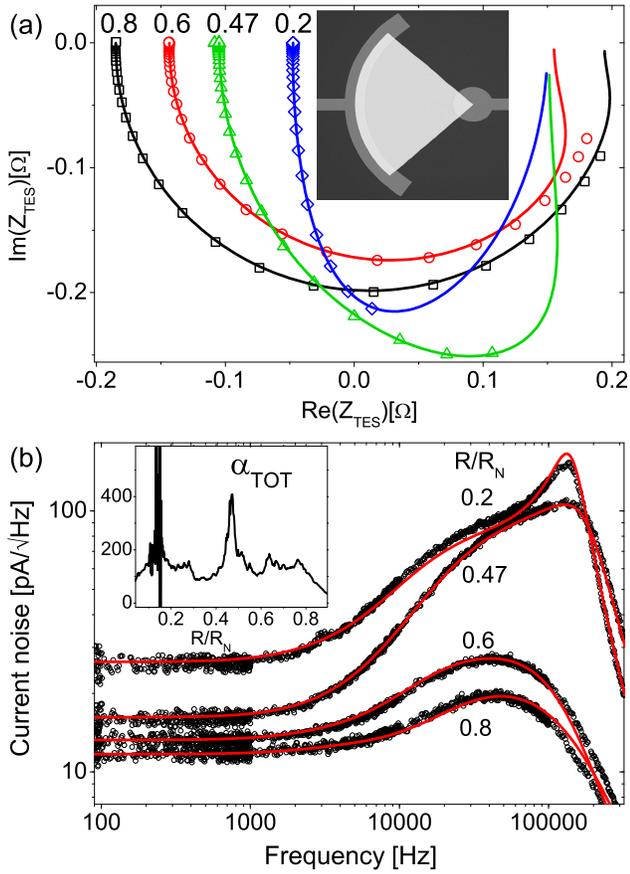}
\caption{(Color online) Measured data (symbols) and fits (lines) for slice TES 2 at $T_{bath}$ = 60 mK. Insets: (a) SEM image of the TES, (b) Total $ \alpha $ calculated from I-V. \label{111fits}}
\end{figure}

The geometrical design of slice TES sample 2 was identical to sample 1, however the SiN and TES layer thickness are different (Table \ref{TEStable}), resulting also to a higher $T_c$. In contrast to sample 1, the second slice TES featured a more complicated transition with very large peaks in $ \alpha _{TOT,IV} $ as shown in the inset of Fig. \ref{111fits}(b). The high values and large variation in $ \alpha _{TOT,IV} $ produces complicated impedance features and high noise levels. Sample 2 was also measured at four different bath temperatures: 140 mK, 110 mK, 85 mK and 60 mK. Fig. \ref{111fits} shows the data for a few representative bias points measured at a bath temperature of 60 mK. The fits were done with the full IH model using Case 2, keeping $C_{TES}$ free.  Now Case 3 is again out of the question due to the complex shapes of the $Z$-data. Case 1 was tried but did not produce as good fits as Case 2.

\begin{figure}
\includegraphics{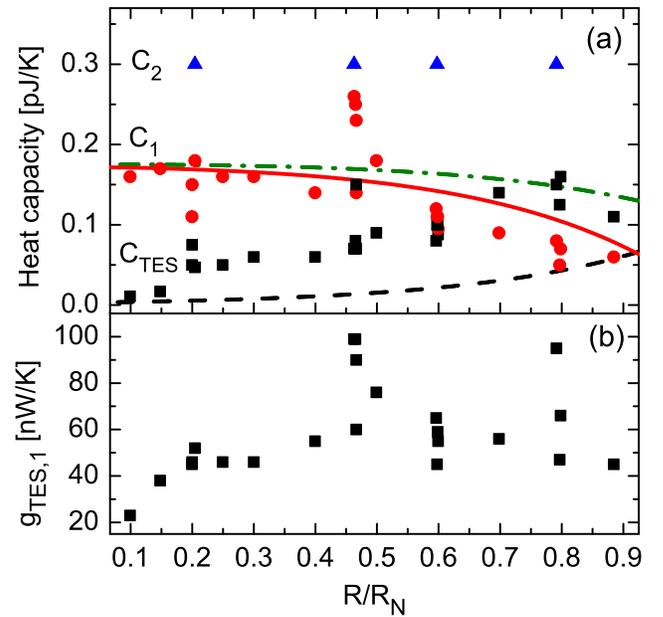}
\caption{(Color online) Parameters obtained from the fits for sample 2 at all values of $T_{bath}$, except $C_{2}$ is shown only at $T_{bath}=$ 60 mK. (a) Squares: $ C_{TES}$, circles: $C_{1} $ and triangles: $ C_{2} $. Dashed black and solid red lines are theoretical calculations for $ C_{TES}$ and $ C_{1}$, respectively, whereas the dash-dotted line is their sum. The reduced BCS value \cite{kozorezov} was used for $ C_{1} $. \label{111param}}
\end{figure}

Again, the main features of both $Z$ and noise data are captured with the IH model. However, at bias points where $\alpha$ is very large (for example $R/R_{N} =0.47 $) it was more difficult to find good fits, and determination of $ Z_{\infty} $ is more uncertain. The inevitable result of this is more scatter in the fitted parameters. In Fig. \ref{111param} we have plotted all the results from all four bath temperature runs in one graph, as $C_{TES}$, $ C_{1} $ and $ g_{TES,1} $ should not depend on $ T_{bath} $ at all according to the thermal model ($C_{2}$ depends on $T_{bath}$ through its influence on $T_{2}$, thus we show only $T_{bath}=$ 60 mK fits). Comparing to Sample 1, we see that $C_{TES}$ and $C_{1}$ have the same trends, and only slightly higher values, consistent with the increase of $T_c$. On the other hand, $C_{2}$ is three times larger. This again supports the picture that $C_{2}$ originates from the SiN membrane, as it is 2.5 times thicker for Sample 2. $ g_{TES,1} $ is of the same order of magnitude, and does not follow a monotonous trend throughout the whole transition. Most of the noise data did not require any additional $M$-noise, all  the fits shown in Fig. \ref{111fits} are without the $M$ parameter.
This, we speculate, could be because of the much higher $\alpha$ compared to Sample 1: The themal noise grows so much that is swamps any possible additional noise sources. 

The data for $ g_{TES,2} $ was not shown because it depends on the bath temperature. For comparison with sample 1, at 60 mK sample 2 had $ g_{TES,2}(T_{TES}) $ = 1.1 nW/K and $ G_{dyn} $ = 0.97 nW/K at 0.5 $ R/R_N $.

As can be seen for bias point $R/R_{N}=0.2$ in Fig. \ref{111fits}, the measured noise sometimes develops a peak near the high-frequency roll-off, when the TES is biased low in the transition. This indicates electrothermal oscillations and it is predicted by our thermal model, as shown by the fit. It arises due to a resonant-like interaction between the electrical and thermal circuits: When $ C_{1} $ starts to decouple from a TES that has a very small heat capacity and large enough $ \alpha $, the TES response becomes oscillatory. 
The increased responsivity means that all noise sources develop a peak, including the  Johnson noise, for example.

\subsection{Square TES}

\begin{figure}
\includegraphics{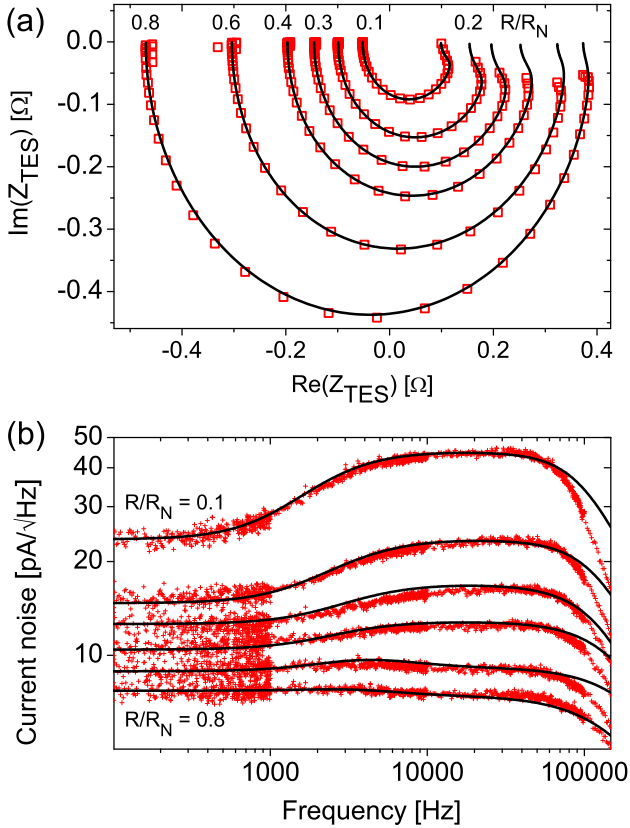}
\caption{(Color online) Measured square TES data (symbols) and fits with IH model (lines). \label{TTESfits}}
\end{figure}

Finally, we show also results for a simple square TES. The measured detector was a bare $\mathrm{300 \mu m} \times \mathrm{300 \mu m}$ TES with Nb bias lines and no extra features added (more sample parameter details in Table \ref{TEStable}). It has a smoothly changing $\alpha_{TOT,IV}$ with a fairly low maximum value of 50. In Fig. \ref{TTESfits} we plot the measured and fitted impedance and noise data. Just as for all the other geometries, our square TES clearly has a large high frequency noise component. Again, the shapes of the impedance curves shows that the model without a hanging block (case 3 discussed in section \ref{sec:slice1}) will not work here either. The IH model again produces a very good fit for both $Z$ and noise, which gives us confidence that the noise in the square TES can be explained by the same mechanism as in the other geometries. Note that no $M$ parameter was needed in these fits.

\begin{figure}
\includegraphics{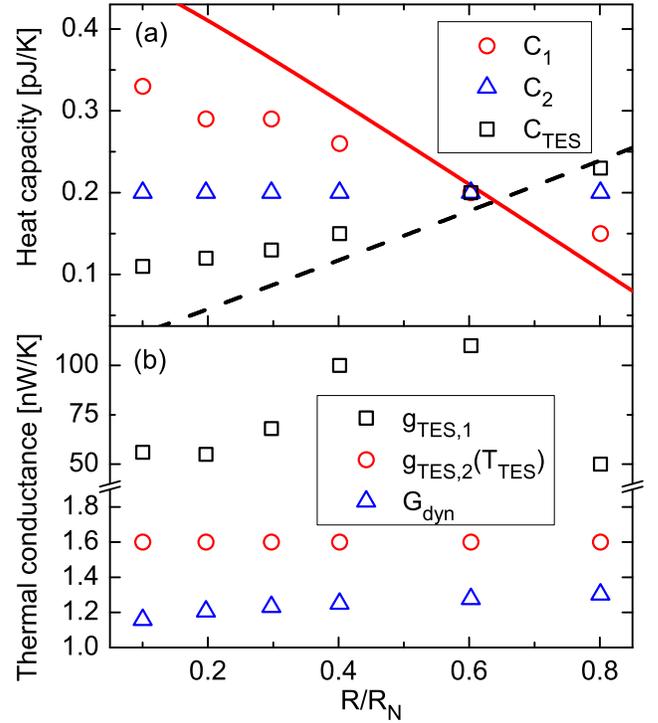}
\caption{(Color online) Parameters from the fits in Fig. \ref{TTESfits}. Lines show the simple estimates for $C_{TES}$ (dashed) and $C_{1}$ (solid) discussed in the text. \label{TTESparam}}
\end{figure}

 Fig. \ref{TTESparam} shows the relevant parameters from the fits. We again observe a decreasing $C_{TES}$ and an increasing $C_{1}$, consistent with our model of thermal decoupling of the N and S regions, and comparable values to the Corbino-like samples. To calculate some estimate for the N and S phase heat capacities, we take a crude model of a linear transition, where the ratio of the volumes is given directly by $R/R_N$, so for example in the middle of the transition we would have exactly half of the device in the superconducting state. This model produces a correct order of magnitude for $C_{TES}$ and $C_{1}$, but naturally no real agreement, although the parameter values do seem to change linearly with $R$, as our overly simple model predicts.  The values of thermal conductances are again consistent with the Corbino devices, and $g_{TES,1}$ does not show any simple dependence on $R$.
  
\section{Conclusions and discussion}

We have fabricated several different geometrical designs of TES devices, and studied their impedance and noise properties. All data are consistent with the picture that a three block thermal circuit is required to accurately describe the data. In all devices, the fitted heat capacity values follow trends which suggest that one of the extra heat capacity blocks arises from the insulating films such as the SiN membrane and, if present, the AlOx layer used in some devices. The other two blocks seem to represent the normal and superconducting regions of the TES film itself, with the observation that the finite value of thermal conductance within the TES film (or a possible thermal boundary resistance between the N and S phases), in combination with the large value of the heat capacity in the superconducting phase near $T_c$, leads to fairly large observable features in the impedance and noise data at high frequencies. In most cases, this internal thermal fluctuation noise component explains all of the "excess noise" in the devices. However, in some devices indications of additional noise sources remained, especially low in the transition. 
  
In our Ti/Au devices the resistivity of the TES film is such that $ g_{TES,1} $ is "low enough". By this we mean that the resulting cut-off frequency of the ITFN noise due to the hanging block is near the electrical cut-off of the readout, so that we can observe its effect in the measured data. Our data, therefore, agrees with the previous suggestion that a large ITFN noise exists in Ti/Au TESs \cite{hoevers,SRON}.

In contrast, the lower resistivity Mo/Au or Mo/Cu devices usually have a lower excess noise level that is white voltage noise, in other words noise that looks like excess Johnson noise \cite{ullom,iyomoto}. 
Our model could perhaps explain part of this difference between Ti based and Mo based devices through the higher thermal conductance $g_{TES,1}$ of the Mo/Au and Mo/Cu TES films. If $g_{TES,1}$ grows, the ITFN noise level will fall, and the cut-off is pushed to a higher frequency.  Thus if the ITFN roll-off happens much after the readout cut-off, ITFN noise will look like excess Johnson noise in the frequency range of the measurement. More measurements on Mo devices are required to test this hypothesis.

The results reported here indicate that in an optimal TES design, the internal thermal conductances within the metallic parts of the device (TES film, absorber) should be maximized. As the actual phase separation between the superconducting and normal regions may be impossible to prevent, the effect of the superconducting heat capacity is minimized if the heat capacity of the normal part of the TES is larger.  We note that the leading designs in terms of reported energy resolution \cite{ullom,iyomoto} employ added normal metal features, and thus we speculate that the reduction of their noise could originate, at least partly, due to the increased normal region heat capacity. 

\begin{acknowledgments}

We acknowledge helpful discussions about data analysis with M. Lindeman, J. van der Kuur, J. Ullom and D. Swetz. This work was supported by the Finnish Funding Agency for Technology and Innovation TEKES and EU through the regional funds, and the Finnish Academy project no. 128532. M. P. would like to thank the National Graduate School in Materials Physics for funding.

\end{acknowledgments}


\end{document}